\documentclass[conference, 9pt]{IEEEtran}
\usepackage{setspace}
\def\ninept{\def\baselinestretch{1.03}\let\normalsize\small\normalsize}
\usepackage{graphicx}
\usepackage{epstopdf}
\IEEEoverridecommandlockouts
\usepackage{hyperref}
\usepackage{algorithm}
\usepackage{algpseudocode}

\usepackage{booktabs, multirow}
\usepackage[table]{xcolor}

\usepackage{amsmath,amsfonts,bm}

\def\eqref#1{equation~\ref{#1}}

\def\1{\bm{1}}

\def\vc{{\bm{c}}}
\def\vd{{\bm{d}}}

\def\vn{{\bm{n}}}

\def\vq{{\bm{q}}}

\def\vv{{\bm{v}}}

\def\vx{{\bm{x}}}
\def\vy{{\bm{y}}}

\def\mE{{\bm{E}}}

\def\mI{{\bm{I}}}

\DeclareMathAlphabet{\mathsfit}{\encodingdefault}{\sfdefault}{m}{sl}
\SetMathAlphabet{\mathsfit}{bold}{\encodingdefault}{\sfdefault}{bx}{n}

\def\sR{{\mathbb{R}}}

\usepackage{cite}
\usepackage{amsmath,amssymb,amsfonts}
\usepackage{graphicx}
\usepackage{textcomp}
\def\BibTeX{{\rm B\kern-.05em{\sc i\kern-.025em b}\kern-.08em
    T\kern-.1667em\lower.7ex\hbox{E}\kern-.125emX}}
\begin{document}
\title{Language-Queried Target Sound Extraction Without Parallel Training Data

\thanks{
This work was supported in part by the National Key Research and Development Program of China under Grant 2022YFC3302800 and the Innovation and Development Joint Funds of Shandong Natural Science Foundation under Grant ZR2022LZH012. The work by Mingjie Shao was supported in part by the National Natural Science Foundation of China under Grant 62401340 and the Natural Science Foundation of Shandong Province under Grant ZR2023QF103. \textit{(Corresponding authors: Ju Liu and Mingjie Shao.)}
}
}

\author{
\IEEEauthorblockN{Hao Ma$^1$, Zhiyuan Peng$^2$, Xu Li$^3$, 
Yukai Li$^1$,
Mingjie Shao$^4$, 
Qiuqiang Kong$^5$,
and Ju Liu$^1$}
\\
\vspace{-2mm}
\IEEEauthorblockA{$^1$School of Information Science and Engineering, Shandong University, Qingdao, China\\
$^2$Department of Computer Science, North Carolina State University, North Carolina, USA  \qquad
$^3$ARC Lab, Tencent PCG\\
$^4$Key Laboratory of System and Control, AMSS, Chinese Academy of Sciences, Beijing, China\\
$^5$The Chinese University of Hong Kong, Hong
Kong, SAR, China}
\\
}

\maketitle

\begin{abstract}
Language-queried target sound extraction (TSE) aims to extract specific sounds from mixtures based on language queries. Traditional fully-supervised training schemes require extensively annotated parallel audio-text data, which are labor-intensive. We introduce a parallel-data-free training scheme, requiring only unlabelled audio clips for TSE model training by utilizing the contrastive language-audio pre-trained model (CLAP). In a vanilla parallel-data-free training stage, target audio is encoded using the pre-trained CLAP audio encoder to form a condition embedding, while during testing, user language queries are encoded by CLAP text encoder as the condition embedding. This vanilla approach assumes perfect alignment between text and audio embeddings, which is unrealistic. Two major challenges arise from training-testing mismatch: the persistent modality gap between text and audio and the risk of overfitting due to the exposure of rich acoustic details in target audio embedding during training. To address this, we propose a retrieval-augmented strategy. Specifically, we create an embedding cache using audio captions generated by a large language model (LLM). During training, target audio embeddings retrieve text embeddings from this cache to use as condition embeddings, ensuring consistent modalities between training and testing and eliminating information leakage. Extensive experiment results show that our retrieval-augmented approach achieves consistent and notable performance improvements over existing state-of-the-art with better generalizability.
\end{abstract}

%
\section{Introduction}
\label{sec:intro}

Humans possess a remarkable ability to focus on specific sounds in noisy environments, a phenomenon known as the \textit{cocktail party effect}. In signal processing, sound separation \cite{SS, MSS} has been extensively researched to address this challenge. Early work focused on domain-specific signals like speech \cite{convtasnet, sepformer} or music \cite{mss_sample, mss_decouple}, but recent advances in deep learning have led to universal sound separation (USS) \cite{kavalerov2019universal, USS_main}, with the ambition to generalize to separate all kinds of sounds. A key development in USS is framing it as query-oriented target sound extraction (TSE) \cite{Label_queried_1, Label_queried_Waveformer, Label_audio_query, language_audio_queried, audio_query, tzinis2023optimal, image_query1, audiosep, dong2023clipsep, ma2024clapsep}, where auxiliary information such as language are introduced to specify what sound to extract.

Despite significant progress in language-queried TSE, practical challenges persist. Current language-queried TSE systems~\cite{LASS, audiosep, ma2024clapsep} heavily depend on large-scale text annotations of sound events for training. This fully-supervised approach requires extensive parallel audio-text datasets to enhance model accuracy and generalizability. However, the high labor cost of annotations limits the availability of parallel data, making it insufficient for scalable universal sound extraction. This scarcity, though under-explored in TSE, is a known challenge in other cross-modal tasks like text-to-image~\cite{NI_1}, text-to-audio~\cite{liu2023audioldm}, and audio captioning~\cite{NI_3}. These studies \cite{NI_1, liu2023audioldm, NI_3} address the issue by leveraging modality-aligned embedding spaces from contrastive cross-modal pre-trained models that are previously trained on massive cross-modal data pairs, eliminating the need for extensive parallel data for specific downstream tasks.

Building on this paradigm, the contrastive language-audio pre-trained model (CLAP)~\cite{CLAP} can be used to eliminate the need for text annotations in TSE. Typically, TSE models are trained using a target audio-text pair mixed with non-target audio, learning to extract the target sound in the audio mixture based on the text query.
To bypass text annotations, CLAP can be cascaded with the TSE model. During training, the target audio is encoded into an audio embedding by the CLAP audio encoder, which the TSE model uses to extract the target sound. During testing, a text query is encoded by the CLAP text encoder into a text embedding for sound extraction.
However, this vanilla approach assumes perfect alignment between text and audio embeddings, which is unrealistic. Two major challenges arise from training-testing mismatch: the persistent modality gap \cite{C3} between text and audio and the risk of overfitting due to the exposure of rich acoustic details in target audio embedding during training.

\begin{figure}
    \centering
    \includegraphics[width=8cm]{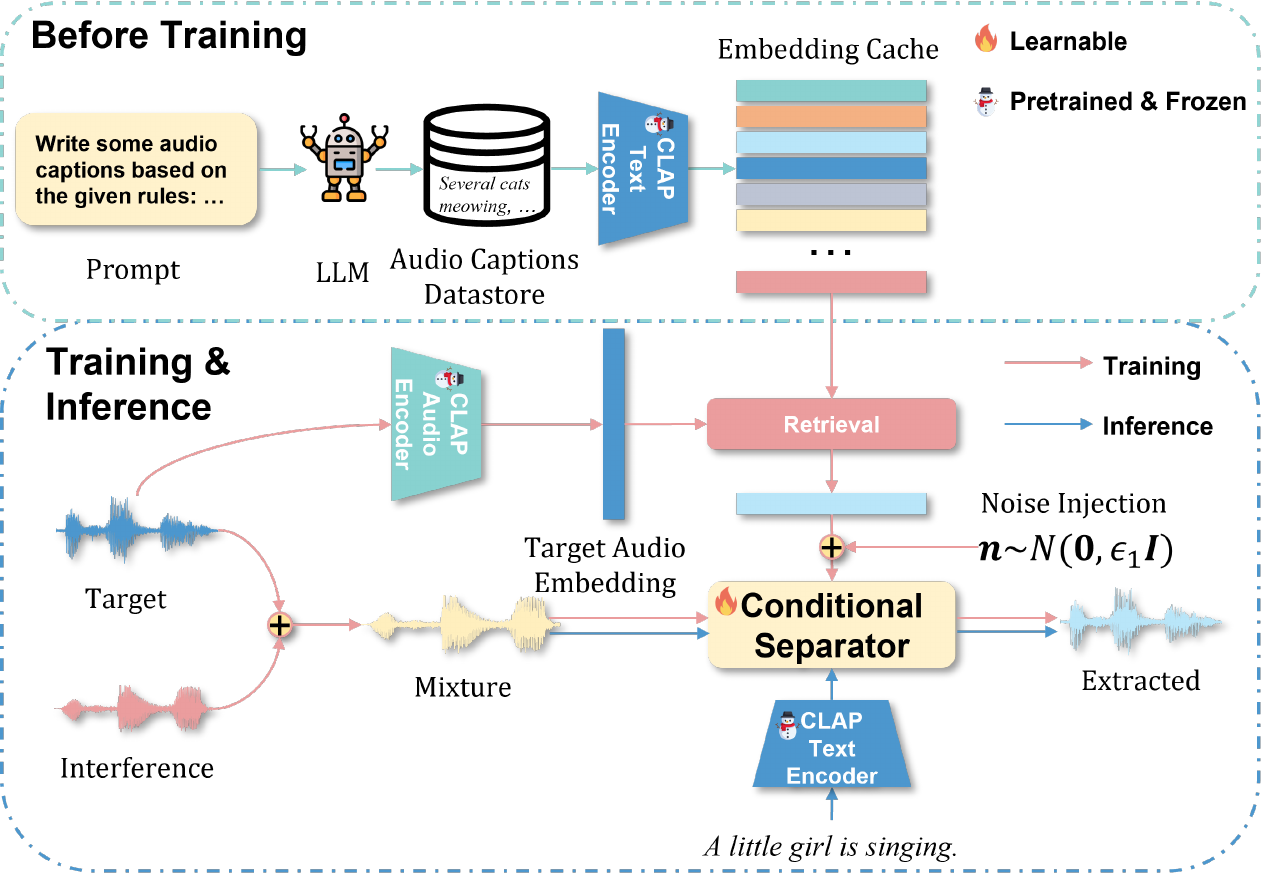}
    \vspace{-2mm}
    \caption{Overview of our proposed retrieval-augmented parallel-data-free training scheme for language-queried target sound extraction.}
    \vspace{-5mm}
    \label{fig:main}
\end{figure}

To address these challenges, we propose a retrieval-augmented, parallel-data-free training paradigm for language-queried TSE models as detailed in Fig. \ref{fig:main}. Before training, we use a large language model to generate diverse audio captions, which are encoded into text embeddings by the pre-trained CLAP text encoder and stored in an embedding cache. During training, we retrieve the most similar text embeddings based on the target audio embeddings and use them as the condition embeddings, guiding the TSE model in sound extraction.
Using this retrieval-augmented strategy which retrieves pre-encoded text embedding for model training, we achieve direct alignment of query modalities between training and testing. Additionally, our approach effectively mitigates the leakage of fine-grained acoustic information present in target audio embeddings—information that query language cannot encapsulate during testing.
We also investigate applying noise injection~\cite{NI_1, NI_2, NI_3} to the condition embedding as an augmentation strategy, enhancing model generalizability.
Extensive experiments show that our retrieval-augmented approach significantly improves performance and generalizability compared to existing methods. By relaxing the need for parallel data, our method scales easily for large-scale training and outperforms previous fully-supervised training schemes across multiple benchmarks, achieving a 1-2 dB improvement in signal-to-distortion ratio (SDR).

\section{Language-queried Target Sound Extraction}
In this section, we propose to address the challenging task of language-queried target sound extraction. We first present the traditional fully-supervised training scheme when parallel audio-text data is available. Then, we introduce the proposed parallel-data-free training scheme with no need for parallel audio-text data.
\label{sec:form&approach}

\subsection{Fully-Supervised Training with Parallel Data}

In fully-supervised training, we have access to audio-text pairs $\vd=\{\vx, \vy\}$, where $\vx$ is an audio clip and $\vy$ is its annotation text. In the mix-and-separate training pipeline, the audio mixture is constructed by sampling several audio clips, treating one of them as the target source and the others as noise, and then summing them as:
\begin{equation}
    \tilde{\vx} = \vx + \vv,
\end{equation}
where $\tilde{\vx} \in \sR^{N}$ denotes the sound mixture of length $N$, $\vx \in \sR^{N}$ denotes the target sound source, and $\vv \in \sR^{N}$ denotes other interfering components.
A query-conditioned target sound extraction model $\mathcal{F}(\cdot)$ parameterized by $\theta$ learns a map from the sound mixture to the target source conditioned on the target source information as:
\begin{equation}
    \hat{\vx} = \mathcal{F}(\tilde{\vx}, \vc; \theta),
\end{equation}
where $\hat{\vx} \in \sR^N$ represents the predicted target sound source and $\vc \in \sR^D$ denotes the $D$-dimensional condition embedding, which is acquired by encoding the annotated text of the target sound by the CLAP text encoder both in the training and testing stages as:
\begin{equation}
    \vc = \text{CLAP}_{\text{text}}(\vy).
\end{equation}
\subsection{Proposed Approach}
In parallel-data-free training, we only have access to audio clips without text annotations. The critical challenge is \textit{how to construct the condition embedding $\vc$ without access to text annotations?} Our methods take advantage of the modality-aligned embedding space of contrastive language-audio pre-trained models. In the following part of this section, we first present a vanilla parallel-data-free training scheme and the widely adopted Gaussian noise injection as an augmentation strategy. We then propose a retrieval-augmented strategy to further mitigate specific challenges posed by existing methods to achieve improved performance.
\subsubsection{Vanilla Parallel-Data-Free Training}

In vanilla parallel-data-free training, the condition embedding $\vc$ is constructed by encoding the target audio $\vx$ with the pre-trained CLAP audio encoder as:
\begin{equation}
    \vc = \text{CLAP}_{\text{audio}}(\vx).
\end{equation}

During testing, we use the pre-trained CLAP text encoder to encode the user language query to control the TSE model. This vanilla strategy works thanks to the modality-aligned embedding space generated by pre-trained CLAP, but it also suffers from performance degradation mainly from the mismatch between training and testing: one is the modality gap arose from the imperfect alignment of audio and text embeddings~\cite{liang2022mind, C3}; the other is the leakage of fine-grained acoustic details in target audio embedding, which cannot be captured by query language, making the model prone to overfitting.

\subsubsection{Gaussian Noise Injection}

Gaussian noise injection has been proven to be effective in bridging the modality gap in many other cross-modal tasks \cite{NI_1, NI_2, NI_3}. This method models the gap between the audio and text embeddings extracted by pre-trained CLAP encoders on parallel audio and text as Gaussian noise with mean $0$ and variance $\epsilon$. Based on the above modeling, the Gaussian noise is used to perturb the condition embeddings extracted from the target audio during parallel-data-free training of a TSE model as:
\begin{equation}
    \vc = \text{CLAP}_{\text{audio}}(\vx) + \vn,
\end{equation}
where $\vn \in \sR^D$ is sampled from a Gaussian distribution $N(\mathbf{0}, \epsilon\mI)$. The noise variance $\epsilon$ is an important hyperparameter in training, which will be explored in Section \ref{sec:pe}.

\subsubsection{Retrieval-Augmented Strategy}
\begin{algorithm}[t]
    \footnotesize
    \caption{retrieval-augmented parallel-data-free training scheme.}
    \label{alg:rae}
    \begin{algorithmic}[1]
        \Require Audio dataset with $K$ audio samples $\mathcal{X}=\{\vx_1, \vx_2,...,\vx_K\}$, embedding cache $\mE \in \mathbb{R}^{M \times D}$, randomly initialized TSE model $\mathcal{F}(\tilde{\vx}, \vc;\theta)$
        \Ensure Optimized model parameters $\theta^*$
        \State Initialize model parameters $\theta$
        \Repeat
            \State Sample a mini-batch of audio: $\mathcal{X}_i \leftarrow \mathrm{sample}(\mathcal{X}) \in \mathbb{R}^{B \times N}$;
            \State Create audio mixture: $\mathcal{\tilde{X}}_i \leftarrow \mathcal{X}_i + \mathrm{shuffle}(\mathcal{X}_i) \in \mathbb{R}^{B \times N}$;
            \State Extract target audio embedding: $\vq \leftarrow \text{CLAP}_{\text{audio}}(\mathcal{X}_i) \in \mathbb{R}^{B \times D}$;
            \State \textcolor{blue}{// Retrieve from the embedding cache}
            \State Compute cosine similarity: $\mathrm{sim} \leftarrow \frac{\vq \mE^T}{||\vq||||\mE||} \in \mathbb{R}^{B \times M}$;
            \State Find indices of maximum similarity: 
            \State $\mathrm{indices} \leftarrow \mathrm{argmax}(\mathrm{sim}, \text{axis}=1) \in \mathbb{R}^{B}$;
            \State Retrieve text embeddings: $\vc \leftarrow \mE[\mathrm{indices}, :] \in \mathbb{R}^{B \times D}$;
            \State Noise injection: $\vc \leftarrow \vc + \vn_1 \sim N(\mathbf{0}, \epsilon_1\mI) \in \mathbb{R}^{B \times D}$;
            \State Compute the gradient: $\nabla_\theta \mathcal{L}(\mathcal{X}_i, \mathcal{F}(\mathcal{\tilde{X}}_i,\vc;\theta))$;
            \State Update model parameters $\theta$ based on the computed gradient;
        \Until{model convergence}
        \State \Return optimized $\theta^* \leftarrow \theta$
    \end{algorithmic}
\end{algorithm}

\begin{table*}[t]
	\caption{Language-queried TSE performance evaluation. ``bal.'' denotes AudioSet \texttt{balanced\_train} and ``unbal.'' denotes AudioSet \texttt{unbalanced\_train}. The term ``Vanilla'' denotes target audio embeddings w/o any augmentation as condition embeddings for training and ``NI'' is noise injection for short. Except for \textit{LASS} and \textit{AudioSep}, all the methods listed in this table are built upon \textit{CLAPSep} \cite{ma2024clapsep}. \dag: reproduction in \cite{ma2024clapsep}, \ddag: ``others'' in \cite{audiosep} is comprised of AudioCaps, Clotho v2, WavCaps, VGGSound.} 
	\centering
        \footnotesize
	\label{tab:1}
        \setlength{\tabcolsep}{0.8mm}{
	\begin{tabular}{lcc||cc|cc|cc|cc|cc||cc}
		\toprule
		\multirow{2}*{\textbf{Method}}&\multirow{1}*{\textbf{Parallel}}&\multirow{2}*{\textbf{Dataset}}&\multicolumn{2}{c}{\textbf{AudioCaps}}&\multicolumn{2}{c}{\textbf{Clotho v2}}&\multicolumn{2}{c}{\textbf{AudioSet}}&\multicolumn{2}{c}{\textbf{MUSIC21}}&\multicolumn{2}{c||}{\textbf{ESC50}}&\multicolumn{2}{c}{\textbf{\textit{Avg}}}\\
		\cmidrule(lr){4-15}
        &\textbf{Data}&&SDRi&SI\text{-}SDRi&SDRi&SI\text{-}SDRi&SDRi&SI\text{-}SDRi&SDRi&SI\text{-}SDRi&SDRi&SI\text{-}SDRi&SDRi&SI\text{-}SDRi\\
        \midrule
        LASS \cite{LASS}&Y&AudioCaps$^\dag$&7.37&6.06&5.93&3.01&5.00&2.02&2.13&-3.64&8.32&6.00&5.75&2.69\\
        AudioSep \cite{audiosep}&Y&unbal.+others$^\ddag$&8.12&7.03&7.26&5.64&8.44&\textbf{7.26}&9.22&7.85&10.23&9.21&8.65&7.40\\
        CLAPSep \cite{ma2024clapsep}&Y&AudioCaps&9.69&8.78&8.63&6.80&\textbf{8.52}&6.88&5.86&1.90&10.88&9.23&8.72&6.72\\
        \midrule
		weakly-supervised&Y&bal.&7.76&6.33&6.95&3.93&8.13&6.69&7.65&5.91&10.55&9.23&8.21&6.42\\
        { -w/ NI}&Y&bal.&7.92&6.58&7.81&5.82&8.04&6.67&7.77&6.27&10.79&9.87&8.47&7.04\\
        \midrule
        \rowcolor{gray!10}Vanilla&N&bal.&7.33&5.81&6.57&4.65&4.33&-0.13&5.31&2.15&8.20&6.15&6.35&3.73\\
        \rowcolor{gray!10}{ -w/ NI}&N&bal.&8.65&7.62&8.38&6.85&6.63&3.99&8.02&6.29&10.99&10.15&8.53&6.98\\
        \rowcolor{green!10}Retrieval &N&bal.&8.76&7.73&8.40&6.73&6.78&4.11&8.23&6.59&11.26&10.40&8.69&7.11\\
        \rowcolor{green!10}{ -w/ NI} &N&bal.&8.70&7.67&8.60&7.14&7.17&4.87&8.28&6.78&11.36&10.63&8.82&7.42\\
        \midrule
        \rowcolor{gray!25}Vanilla w/ NI &N&$\mathcal{DS}_{large}$&9.60&8.71&9.06&7.59&7.11&4.20&9.85&8.37&12.20&11.40&9.56&8.05\\
        \rowcolor{green!25}Retrieval w/ NI&N&$\mathcal{DS}_{large}$&\textbf{9.75}&\textbf{8.92}&\textbf{9.43}&\textbf{8.12}&8.09&5.75&\textbf{10.24}&\textbf{9.11}&\textbf{12.55}&\textbf{11.89}&\textbf{10.00}&\textbf{8.76}\\
		\bottomrule
	\end{tabular}}
 \vspace{-2mm}
\end{table*}

Although many previous works \cite{NI_1, NI_2, NI_3} have demonstrated that Gaussian noise injection effectively addresses the mismatch issue for cross-modal tasks, there still exists a performance gap compared to training with manually labeled data pairs \cite{NI_3}. To further bridge this gap, we introduce a retrieval-augmented strategy that directly aligns query modalities between training and testing while addressing the leakage of acoustic information that language-based queries cannot encapsulate. As shown in Fig. \ref{fig:main}, before the model training, we first generate massive audio captions by prompting a large language model following \cite{mei2023wavcaps}. These audio captions are then encoded by the pre-trained CLAP text encoder to create an embedding cache $\mE \in \sR^{M \times D}$, where $M$ is the number of audio captions generated by LLM.
In training, the target audio $\vx$ is encoded by the CLAP audio encoder as a query to retrieve the most similar text embedding in the embedding cache $\mE$. The retrieved text embedding is considered as a condition embedding $\vc$ and further augmented by a Gaussian perturbation as:
\begin{equation}
    \vc = \mathrm{Retrieve}(\mE, \text{query}=\text{CLAP}_{\text{audio}}(\vx)) + \vn_1,
\end{equation}
where we retrieve the most similar text embedding in the embedding cache according to cosine similarity; and $\vn_1 \sim N(\mathbf{0}, \epsilon_1\mI) \in \sR^D$. We explore the setting of $\epsilon_1$ in Section \ref{sec:pe}. A more detailed procedure for retrieval-augmented parallel-data-free training of a language-queried target sound extraction model is given in Alg. \ref{alg:rae}.

\section{Experiments}

\subsection{Datasets and Evaluation Metrics}

We consider two data sources commonly used in the literature \cite{audiosep, ma2024clapsep, dong2023clipsep, USS_main} for model training: AudioSet \cite{audioset} and FreeSound \cite{font2013freesound}. AudioSet is a large-scale audio collection from YouTube videos, while FreeSound is an online collaborative sound-sharing site. We collect a total of 1,912,173 audio clips from the \texttt{unbalanced\_train} split of AudioSet and 262,300 audio clips from FreeSound. All audio clips are sampled at 32kHz.

We first perform quick preliminary experiments on the \texttt{balanced\_train} split of AudioSet, which contains 20,550 audio clips, to evaluate the effectiveness of our proposed method and determine the hyperparameter settings. We then scale up the training data to include all collected audio clips from AudioSet and FreeSound to further validate the effectiveness of our method on large-scale non-parallel data. We refer to this large dataset as $\mathcal{DS}_{large}$.

We follow recipes in \cite{audiosep, ma2024clapsep} to prepare evaluation datasets. We perform evaluations on AudioCaps \cite{audiocaps}, Clotho v2 \cite{drossos2020clotho}, the AudioSet \texttt{test} split, MUSIC21 \cite{image_query1} and ESC50 \cite{ESC50}, which cover a wide range of natural sound and music to comprehensively validate the effectiveness of our proposed method. Note that in AudioCaps and Clotho v2, the annotations are human-written audio captions in natural language form, while in AudioSet, MUSIC21, and ESC50, the annotations are labels of sound events. In testing, we convert these labels to language-like audio captions by adding the prefix \textit{``The sound of ''}.

We adopt the commonly used SDRi and SI\text{-}SDRi \cite{sdr} as evaluation metrics, which are defined as:
\begin{align}
\label{SDRi}
    \mathrm{SDRi}(\hat{\vx},\tilde{\vx}, \vx) &= \mathrm{SDR}(\hat{\vx},\vx)-\mathrm{SDR}(\tilde{\vx},\vx),\\
    \label{SISDRi}
    \mathrm{SI\text{-}SDRi}(\hat{\vx},\tilde{\vx}, \vx) &= \mathrm{SI\text{-}SDR}(\hat{\vx},\vx)-\mathrm{SI\text{-}SDR}(\tilde{\vx},\vx),
\end{align}
where
\begin{align}
\label{SDR}
    \mathrm{SDR}(\hat{\vx},\vx) &= 10*\mathrm{log}_{10} \left(\frac{||\vx||^2}{||\vx - \hat{\vx}||^2} \right),\\
\label{SISDR}
    \mathrm{SI\text{-}SDR}(\hat{\vx},\vx) &= 10*\mathrm{log}_{10}\left(\frac{||\frac{\hat{\vx}^{\top}\vx}{||\vx||^2}\vx||^2}{||\frac{\hat{\vx}^{\top}\vx}{||\vx||^2}\vx- \hat{\vx}||^2}\right).
\end{align}
\subsection{Implementation Details}
We build the parallel-data-free training framework upon the state-of-the-art (SOTA) language-queried target sound extraction model, CLAPSep \cite{ma2024clapsep}, which is a transformer-based model. For experiments on the \texttt{balanced\_train} split, we train the model on a single RTX 3090 GPU with a batch size of 32 for a total of 96,450 steps. The initial learning rate is set to 1e-4, which decays exponentially by a factor of 0.32 at steps 31,150 and 64,300. For experiments on $\mathcal{DS}_{large}$, we train the model on four RTX 3090 GPUs with a global batch size of 64 for a total of 509,400 steps. The initial learning rate is set to 2e-4, which decays exponentially by a factor of 0.32 at steps 169,800 and 339,600. We adopt the negative SI\text{-}SDR between the model estimation and the target audio as the loss function as detailed in \eqref{SISDR}.

\subsection{Results and Analysis}

\label{sec:pe}
In this section, we first perform preliminary experiments on AudioSet \texttt{balanced\_train} to demonstrate the validity of the proposed method and to determine the hyperparameter settings. Then we scale up the training data to demonstrate the effectiveness of our proposed method in leveraging large-scale unlabelled audio data for training a language-queried TSE model with strong generalizability.

\subsubsection{Effectiveness of Retrieval-Augmented Strategy}

We show the performance of various parallel-data-free training schemes in Table \ref{tab:1}. As can be seen from these results, the \textit{vanilla} strategy, i.e., the direct use of target audio embedding without any augmentation as the condition embedding for training, performs the worst. This indicates that the mismatch during training and testing time introduced by this strategy causes a significant performance drop. The mismatch problem can be solved effectively by Gaussian noise injection. As can be seen from the results in the table (\textit{Vanilla w/ NI on bal.}), Gaussian noise injection brings a performance improvement of more than 2dB on average for the vanilla strategy. Further, the retrieval-augmented strategy proposed in this paper combined with a certain degree of noise injection (\textit{Retrieval w/ NI on bal.}) performs optimally over all kinds of parallel-data-free training schemes. It is worth noting that the better performance achieved by our proposed method does not rely on Gaussian noise injection, which is shown by the performance gap between ``\textit{Retrieval}'' and ``\textit{Retrieval w/ NI}'' is smaller than that between vanilla strategies.
This indicates that the proposed method has more effectively addressed the training-testing mismatch problem by retrieving pre-encoded text embeddings.
\subsubsection{Parallel-Data-Free Versus Weakly-Supervised}
We conduct weakly-supervised experiments by invoking the label-formed annotations on AudioSet and transforming the audio labels annotated to each of the audio clips into captions by prefixing them with \textit{``The sound of ''}. The detailed results on each of the evaluation benchmarks are shown in Table \ref{tab:1}. Comparing the weakly-supervised training scheme with our proposed parallel-data-free training schemes, the weakly-supervised method suppresses the parallel-data-free methods only on AudioSet but falls behind on other evaluation benchmarks. This is because the query language in the AudioSet \texttt{test} set is defined in the same way as in weakly-supervised training. This phenomenon shows that even when one-to-one weak labels (like those in AudioSet) are available, our method can achieve superior performance and better generalizability on natural language-formed user queries without requiring such annotations.
\subsubsection{Effect of Gaussian Perturbation Strength}
\begin{figure}
    \centering
    \includegraphics[width=8.5cm]{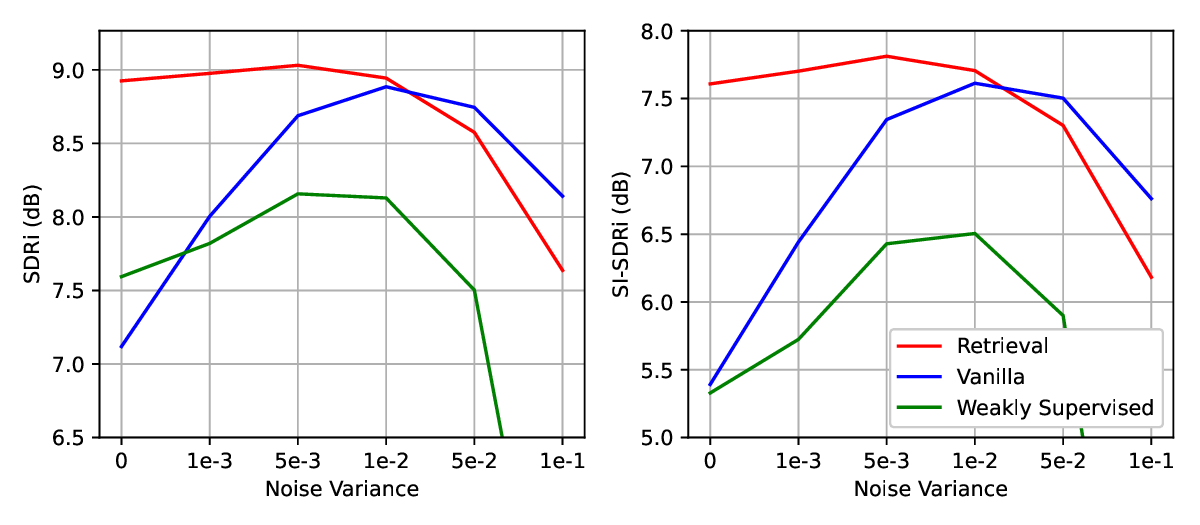}
    \vspace{-0.5mm}
    \caption{Effect of Gaussian perturbation strength.}
    \vspace{-3mm}
    \label{fig:nstd}
\end{figure}

The choice of the Gaussian perturbation variance has a significant effect on the performance of parallel-data-free training \cite{NI_1, NI_2, NI_3}. To determine the Gaussian perturbation strength, we vary the variation of injected Gaussian noise from 0 to 1e-1 and train a TSE model for each value. We run experiments on a held-out validation set consisting of audio-text pairs from AudioCaps and Clotho v2 \texttt{validation} split and present the SDRi and SI\text{-}SDRi in Fig. \ref{fig:nstd}.

The green lines in Fig. \ref{fig:nstd} illustrate the effect of noise variance $\epsilon$ on the performance of the vanilla strategy, which injects Gaussian noise directly into the target audio embedding to form the condition embedding in training. As shown in Fig. \ref{fig:nstd}, the appropriate choice of noise variance has a crucial impact on the effectiveness of this approach. Specifically, when the noise variance is small, the mismatch between training and testing will cause a significant drop in performance. On the other hand, if the noise variance is too large, the useful information in audio embeddings is overwhelmed by too much noise. According to Fig. \ref{fig:nstd}, we set the noise variance to 1e-2 for all subsequent experiments for the vanilla strategy.

The red lines in Fig. \ref{fig:nstd} illustrate the effect of the noise variance $\epsilon_1$ on the performance of our proposed retrieval-augmented strategy. The experimental results show that for our proposed retrieval-augmented strategy, Gaussian noise injection of appropriate intensity also improves performance to some extent, but the effect is far less dramatic than that for the vanilla strategy, further demonstrating that the better performance achieved by the proposed retrieval-augmented strategy does not depend on the noise perturbation since the training-testing mismatch has already been effectively addressed by retrieving pre-encoded text embeddings.
\subsubsection{Effect of LLM-Produced Captions}
\begin{figure}
    \centering
    \includegraphics[width=8.5cm]{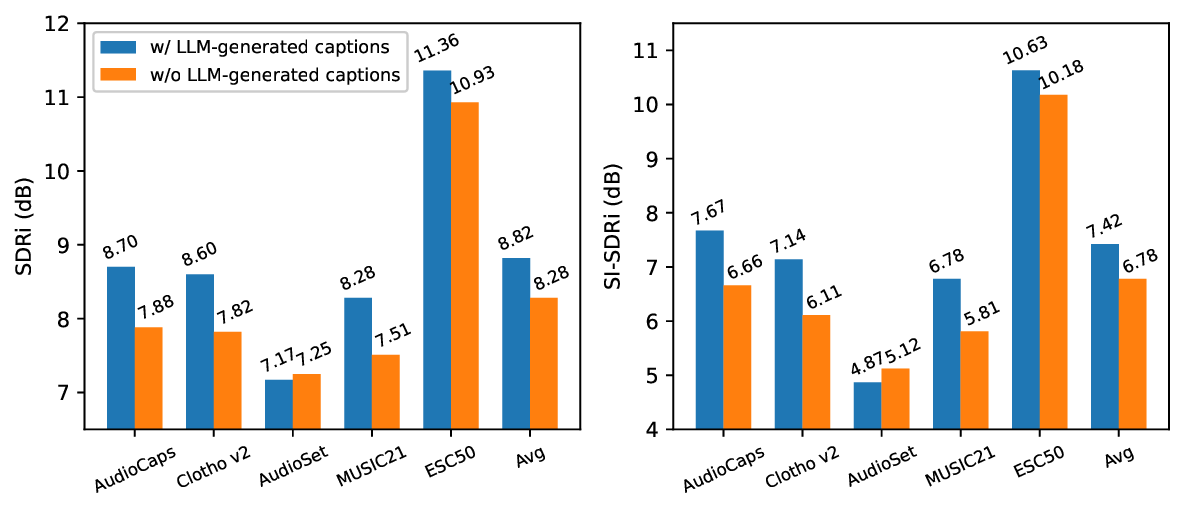}
    \vspace{-1mm}
    \caption{Effect of LLM-produced captions.}
    \vspace{-3mm}
    \label{fig:llm}
\end{figure}

Our proposed retrieval-augmented parallel-data-free training scheme requires retrieving corresponding text embeddings from a pre-encoded text embedding cache to construct condition embeddings. Therefore, the quality of this embedding cache plays a crucial role in the performance of the proposed retrieval strategy. We suggest using LLM-produced audio captions to construct this embedding cache. In this section, we design experiments to verify the impact of LLM-produced audio captions on the performance of our proposed method. Specifically, we first establish a baseline system where we take all the 527 audio labels from AudioSet and construct 527 audio captions by prefixing each with \textit{``The sound of "}. These captions are then encoded into text embeddings using the CLAP text encoder and stored in an embedding cache, thus forming our baseline system. Then we invoke WavCaps \cite{mei2023wavcaps} to simulate LLM-produced audio captions. We encode all the 403,050 LLM-produced audio captions from WavCaps into text embeddings, which are additionally added to the embedding cache in the baseline system. We train a new TSE model for each of the two settings and perform evaluation experiments on all five datasets. The results are shown in Fig. \ref{fig:llm}. The experimental results show that the proposed method, combined with the LLM-produced audio captions, achieves significant performance improvements on most evaluation benchmarks.
\subsubsection{Scale up the Training Data}
In this section, we scale up the training data to include all collected audio clips from AudioSet and FreeSound to further validate the effectiveness of our method in leveraging large-scale non-parallel data. The corresponding results are presented in the last two rows of Table \ref{tab:1}. From these results, we can see that the proposed parallel-data-free training schemes perform better on all evaluation benchmarks after scaling up the training data. These results also outperform the current SOTA-supervised training schemes on most of the evaluated benchmarks, thus demonstrating the effectiveness of the proposed method in efficiently leveraging a large amount of audio data without annotations to train a language-queried target sound extraction model with strong generalizability.

\section{Conclusion}

In this work, we present a novel parallel-data-free training paradigm for language-queried target sound extraction, leveraging the modality-aligned embedding space generated by the pre-trained CLAP model. By harnessing unlabeled audio clips, our model deftly sidesteps the necessity for labor-intensive text-audio pairings. We address challenges like modality gap and information leakage through a retrieval-augmented strategy, employing an embedding cache constructed from LLM-produced audio captions. Our extensive evaluation on several benchmarks demonstrates that this parallel-data-free training approach achieves consistent and notable performance improvements over existing state-of-the-art with better generalizability. This method significantly reduces the reliance on human-annotated audio-text data pairs, offering a more adaptable and scalable solution for TSE.


\bibliographystyle{ieeetr} 
\bibliography{strings,refs}

\begin{thebibliography}{10}

\bibitem{SS}
D.~Wang and J.~Chen, ``Supervised speech separation based on deep learning: An overview,'' {\em IEEE/ACM Trans. Audio, Speech, Lang. Process.}, vol.~26, no.~10, pp.~1702--1726, 2018.

\bibitem{MSS}
Z.~Rafii, A.~Liutkus, F.-R. Stöter, S.~I. Mimilakis, D.~FitzGerald, and B.~Pardo, ``An overview of lead and accompaniment separation in music,'' {\em IEEE/ACM Trans. Audio, Speech, Lang. Process.}, vol.~26, no.~8, pp.~1307--1335, 2018.

\bibitem{convtasnet}
Y.~Luo and N.~Mesgarani, ``{Conv-TasNet}: Surpassing ideal time–frequency magnitude masking for speech separation,'' {\em IEEE/ACM Trans. Audio, Speech, Lang. Process.}, vol.~27, no.~8, pp.~1256--1266, 2019.

\bibitem{sepformer}
C.~Subakan, M.~Ravanelli, S.~Cornell, M.~Bronzi, and J.~Zhong, ``Attention is all you need in speech separation,'' in {\em Proc. IEEE Int. Conf. Acoustics Speech Signal Process. (ICASSP)}, pp.~21--25, 2021.

\bibitem{mss_sample}
Y.~Luo and J.~Yu, ``Music source separation with band-split {RNN},'' {\em IEEE/ACM Trans. Audio, Speech, Lang. Process.}, vol.~31, pp.~1893--1901, 2023.

\bibitem{mss_decouple}
Q.~Kong, Y.~Cao, H.~Liu, K.~Choi, and Y.~Wang, ``Decoupling magnitude and phase estimation with deep {ResUNet} for music source separation.,'' in {\em Proc. Int. Conf. Music Inf. Retr. (ISMIR)}, pp.~342--349, 2021.

\bibitem{kavalerov2019universal}
I.~Kavalerov, S.~Wisdom, H.~Erdogan, B.~Patton, K.~Wilson, J.~Le~Roux, and J.~R. Hershey, ``Universal sound separation,'' in {\em Proc. IEEE Workshop Appl. Signal Process. Audio Acoust. (WASPAA)}, pp.~175--179, IEEE, 2019.

\bibitem{USS_main}
Q.~Kong, K.~Chen, H.~Liu, X.~Du, T.~Berg-Kirkpatrick, S.~Dubnov, and M.~D. Plumbley, ``Universal source separation with weakly labelled data,'' {\em arXiv preprint arXiv:2305.07447}, 2023.

\bibitem{Label_queried_1}
T.~Ochiai, M.~Delcroix, Y.~Koizumi, H.~Ito, K.~Kinoshita, and S.~Araki, ``Listen to what you want: Neural network-based universal sound selector,'' in {\em Proc. INTERSPEECH}, pp.~1441--1445, 2020.

\bibitem{Label_queried_Waveformer}
B.~Veluri, J.~Chan, M.~Itani, T.~Chen, T.~Yoshioka, and S.~Gollakota, ``Real-time target sound extraction,'' in {\em Proc. IEEE Int. Conf. Acoustics Speech Signal Process. (ICASSP)}, 2023.

\bibitem{Label_audio_query}
M.~Delcroix, J.~B. Vázquez, T.~Ochiai, K.~Kinoshita, Y.~Ohishi, and S.~Araki, ``{SoundBeam}: Target sound extraction conditioned on sound-class labels and enrollment clues for increased performance and continuous learning,'' {\em IEEE/ACM Trans. Audio, Speech, Lang. Process.}, vol.~31, pp.~121--136, 2023.

\bibitem{language_audio_queried}
K.~Kilgour, B.~Gfeller, Q.~Huang, A.~Jansen, S.~Wisdom, and M.~Tagliasacchi, ``{Text-Driven Separation of Arbitrary Sounds},'' in {\em Proc. INTERSPEECH}, pp.~5403--5407, 2022.

\bibitem{audio_query}
K.~Chen*, X.~Du*, B.~Zhu, Z.~Ma, T.~Berg-Kirkpatrick, and S.~Dubnov, ``Zero-shot audio source separation via query-based learning from weakly-labeled data,'' in {\em Proc. AAAI Conf. Artif. Intell. (AAAI)}, pp.~4441--4449, 2022.

\bibitem{tzinis2023optimal}
E.~Tzinis, G.~Wichern, P.~Smaragdis, and J.~Le~Roux, ``Optimal condition training for target source separation,'' in {\em Proc. IEEE Int. Conf. Acoustics Speech Signal Process. (ICASSP)}, pp.~1--5, IEEE, 2023.

\bibitem{image_query1}
H.~Zhao, C.~Gan, A.~Rouditchenko, C.~Vondrick, J.~McDermott, and A.~Torralba, ``The sound of pixels,'' in {\em Proc. Eur. Conf. Comput. Vis. (ECCV)}, pp.~570--586, 2018.

\bibitem{audiosep}
X.~Liu, Q.~Kong, Y.~Zhao, H.~Liu, Y.~Yuan, Y.~Liu, R.~Xia, Y.~Wang, M.~D. Plumbley, and W.~Wang, ``Separate anything you describe,'' {\em IEEE/ACM Trans. Audio, Speech, Lang. Process.}, pp.~1--15, 2024.

\bibitem{dong2023clipsep}
H.-W. Dong, N.~Takahashi, Y.~Mitsufuji, J.~McAuley, and T.~Berg-Kirkpatrick, ``{CLIPSep}: Learning text-queried sound separation with noisy unlabeled videos,'' in {\em Proc. Int. Conf. Learn. Represent. (ICLR)}, 2023.

\bibitem{ma2024clapsep}
H.~Ma, Z.~Peng, X.~Li, M.~Shao, X.~Wu, and J.~Liu, ``Clapsep: Leveraging contrastive pre-trained model for multi-modal query-conditioned target sound extraction,'' {\em IEEE/ACM Trans. Audio, Speech, Lang. Process.}, vol.~32, pp.~4945--4960, 2024.

\bibitem{LASS}
X.~Liu, H.~Liu, Q.~Kong, X.~Mei, J.~Zhao, Q.~Huang, M.~D. Plumbley, and W.~Wang, ``Separate what you describe: Language-queried audio source separation,'' in {\em Proc. INTERSPEECH}, pp.~1801--1805, 2022.

\bibitem{NI_1}
Y.~Zhou, R.~Zhang, C.~Chen, C.~Li, C.~Tensmeyer, T.~Yu, J.~Gu, J.~Xu, and T.~Sun, ``Towards language-free training for text-to-image generation,'' in {\em Proc. IEEE/CVF Conf. Comput. Vis. Pattern Recog. (CVPR)}, pp.~17907--17917, 2022.

\bibitem{liu2023audioldm}
H.~Liu, Z.~Chen, Y.~Yuan, X.~Mei, X.~Liu, D.~Mandic, W.~Wang, and M.~D. Plumbley, ``{AudioLDM}: Text-to-audio generation with latent diffusion models,'' {\em Proc. Int. Conf. Mach. Learn. (ICML)}, pp.~21450--21474, 2023.

\bibitem{NI_3}
S.~Deshmukh, B.~Elizalde, D.~Emmanouilidou, B.~Raj, R.~Singh, and H.~Wang, ``Training audio captioning models without audio,'' in {\em Proc. IEEE Int. Conf. Acoustics Speech Signal Process. (ICASSP)}, pp.~371--375, 2024.

\bibitem{CLAP}
Y.~Wu, K.~Chen, T.~Zhang, Y.~Hui, T.~Berg-Kirkpatrick, and S.~Dubnov, ``Large-scale contrastive language-audio pretraining with feature fusion and keyword-to-caption augmentation,'' in {\em Proc. IEEE Int. Conf. Acoustics Speech Signal Process. (ICASSP)}, 2023.

\bibitem{C3}
Y.~Zhang, E.~Sui, and S.~Yeung-Levy, ``Connect, collapse, corrupt: Learning cross-modal tasks with uni-modal data,'' in {\em Proc. Int. Conf. Learn. Represent. (ICLR)}, 2024.

\bibitem{NI_2}
D.~Nukrai, R.~Mokady, and A.~Globerson, ``Text-only training for image captioning using noise-injected {CLIP},'' in {\em Find. Assoc. Comput. Linguist.: EMNLP}, pp.~4055--4063, 2022.

\bibitem{liang2022mind}
V.~W. Liang, Y.~Zhang, Y.~Kwon, S.~Yeung, and J.~Y. Zou, ``Mind the gap: Understanding the modality gap in multi-modal contrastive representation learning,'' {\em Proc. Adv. Neural Inf. Process. Syst. (NeurIPS)}, vol.~35, pp.~17612--17625, 2022.

\bibitem{mei2023wavcaps}
X.~Mei, C.~Meng, H.~Liu, Q.~Kong, T.~Ko, C.~Zhao, M.~D. Plumbley, Y.~Zou, and W.~Wang, ``Wavcaps: A chatgpt-assisted weakly-labelled audio captioning dataset for audio-language multimodal research,'' {\em IEEE/ACM Trans. Audio, Speech, Lang. Process.}, vol.~32, pp.~3339--3354, 2024.

\bibitem{audioset}
J.~F. Gemmeke, D.~P.~W. Ellis, D.~Freedman, A.~Jansen, W.~Lawrence, R.~C. Moore, M.~Plakal, and M.~Ritter, ``{Audio Set}: An ontology and human-labeled dataset for audio events,'' in {\em Proc. IEEE Int. Conf. Acoustics Speech Signal Process. (ICASSP)}, pp.~776--780, 2017.

\bibitem{font2013freesound}
F.~Font, G.~Roma, and X.~Serra, ``Freesound technical demo,'' in {\em Proc. ACM Multimedia Conf. (ACM-MM)}, pp.~411--412, 2013.

\bibitem{audiocaps}
C.~D. Kim, B.~Kim, H.~Lee, and G.~Kim, ``{AudioCaps}: Generating captions for audios in the wild,'' in {\em Proc. Conf. N. Am. Chapter Assoc. Comput. Linguistics: Hum. Lang. Technol. (NAACL-HLT)}, pp.~119--132, 2019.

\bibitem{drossos2020clotho}
K.~Drossos, S.~Lipping, and T.~Virtanen, ``Clotho: An audio captioning dataset,'' in {\em Proc. IEEE Int. Conf. Acoustics Speech Signal Process. (ICASSP)}, pp.~736--740, 2020.

\bibitem{ESC50}
K.~J. Piczak, ``{ESC}: Dataset for environmental sound classification,'' in {\em Proc. ACM Multimedia Conf. (ACM-MM)}, pp.~1015--1018, 2015.

\bibitem{sdr}
J.~L. Roux, S.~Wisdom, H.~Erdogan, and J.~R. Hershey, ``{SDR} – half-baked or well done?,'' in {\em Proc. IEEE Int. Conf. Acoustics Speech Signal Process. (ICASSP)}, pp.~626--630, 2019.

\end{thebibliography}

\end{document}